\documentclass[10pt,twocolumn,twoside,journal]{IEEEtran}
\IEEEoverridecommandlockouts
\usepackage{subfigure} 
\usepackage{graphicx}

\usepackage{amsmath,graphicx,amssymb,mathtools,bm}
\usepackage{subfigure}
\usepackage{hyperref}
\usepackage{cite}
\usepackage{amsmath,amssymb,amsfonts}  
\usepackage{textcomp}
\usepackage{xcolor}

\usepackage{verbatim}  
\usepackage{bm}  
\usepackage{mathrsfs} 
 \usepackage{} 
\usepackage{booktabs}
\usepackage{textcomp} 
\usepackage{multirow}  
\usepackage{lettrine}  
\usepackage{graphicx} 
\usepackage{color} 
\usepackage{amsmath}
\usepackage{amssymb}
\usepackage{stfloats} 
\usepackage{caption}

\usepackage{color}
\usepackage[lined,boxed,commentsnumbered]{algorithm2e}

\begin{document}
	\newcommand{\tabincell}[2]{\begin{tabular}{@{}#1@{}}#2\end{tabular}}
   \newtheorem{Property}{\it Property} 
  
 \newtheorem{Proposition}{\bf Proposition}
\newtheorem{remark}{Remark}
\newenvironment{Proof}{{\indent \it Proof:}}{\hfill $\blacksquare$\par}

\title{Flexible Cylindrical Arrays with Movable Antennas for MISO System: Beamforming and Position Optimization}
\author{Jiahe Guo, Songjie Yang, Xiangyu Dong, Jiapan Yang, Junfeng Deng, \\ Zhongpei Zhang, \IEEEmembership{Member,~IEEE}, and Chau Yuen, \IEEEmembership{Fellow,~IEEE}


\thanks{
    	Jiahe Guo, Songjie Yang, Xiangyu Dong, Junfeng Deng and Zhongpei Zhang are with the National Key Laboratory of Wireless Communications, University of Electronic Science and Technology of China, Chengdu 611731, China. (e-mail: guojiahe@std.uestc.edu.cn; yangsongjie@std.uestc.edu.cn; dongxiangyu@std.uestc.edu.cn; dengjunfeng@std.uestc.edu.cn; zhangzp@uestc.edu.cn). \emph{(Corresponding Author: Zhongpei Zhang)}
    	
		Jiapan Yang is with School of Electronic and Information Engineering, Beijing Jiaotong University, Beijing, China (e-mail: jiapanyang@bjtu.edu.cn).
		
		Chau Yuen is with the School of Electrical and Electronics Engineering, Nanyang Technological University (e-mail: chau.yuen@ntu.edu.sg).}}

\maketitle

\begin{abstract}
As wireless communication advances toward the 6G era, the demand for ultra-reliable, high-speed, and ubiquitous connectivity is driving the exploration of new degrees-of-freedom (DoFs) in communication systems. Among the key enabling technologies, Movable Antennas (MAs) integrated into Flexible Cylindrical Arrays (FCLA) have shown great potential in optimizing wireless communication by providing spatial flexibility. This paper proposes an innovative optimization framework that leverages the dynamic mobility of FCLAs to improve communication rates and overall system performance. 
By employing Fractional Programming (FP) for alternating optimization of beamforming and antenna positions, the system enhances throughput and resource utilization. 
Additionally, a novel Constrained Grid Search-Based Adaptive Moment Estimation Algorithm (CGS-Adam) is introduced to optimize antenna positions while adhering to antenna spacing constraints. Extensive simulations validate that the proposed system, utilizing movable antennas, significantly outperforms traditional fixed antenna optimization, achieving up to a 31\% performance gain in general scenarios. The integration of FCLAs in wireless networks represents a promising solution for future 6G systems, offering improved coverage, energy efficiency, and flexibility.

\end{abstract}
\begin{IEEEkeywords}
Movable Antennas, Flexible Cylindrical Arrays, Fractional Programming, constrained Grid Search-Based Adaptive Moment Estimation Algorithm, beamforming, antenna position, Multi-User MISO.
\end{IEEEkeywords} 
\section{Introduction}


\IEEEPARstart{A}{s} the evolution of wireless communication accelerates towards the 6G era, transformative advancements are being conceptualized to address the ever-growing demand for ultra-reliable, high-speed, and ubiquitous connectivity. 6G networks are anticipated to power revolutionary applications such as holographic telepresence, advanced industrial automation, and comprehensive environmental sensing. These applications require the exploration of new degrees-of-freedom (DoFs) in communication systems. 
To address these challenges, several key enabling technologies have emerged as critical research areas. These include Integrated Sensing and Communication (ISAC)\cite{MA_ISAC2,MA_ISAC_MISO}, Reconfigurable Intelligent Surfaces (RISs)\cite{RIS1, RIS2, RIS3}, 
and leveraging the sparsity inherent in millimeter-wave (mmWave) and terahertz channels for beamspace signal processing\cite{mmw1,mmw2}. 
Additionally, exploring near-field spherical-wave channels to exploit the distance DoF represents a pivotal strategy for advancing next-generation communication systems \cite{NF1,NF2,NF3}. These innovations contribute to enhancing the efficiency, flexibility, and performance of modern communication networks, and are poised to unlock the full potential of wireless channels. 
Despite these significant advancements, exploring DoFs in communication systems remains a highly promising avenue for further research, as it offers potential for enhancing system capacity, improving data rates, and optimizing resource allocation in increasingly complex wireless environments.

Recently, Movable Antennas (MAs) have revolutionized wireless communication by introducing spatial flexibility. Unlike fixed-position array antennas, MAs, connected to the radio frequency (RF) chain via flexible cables, can be dynamically repositioned in real time using actuators. This adaptability enables optimization of channel conditions, reducing interference, fading, and enhancing signal strength and overall communication quality \cite{MA4_zon}. These advantages make MAs ideal for maintaining stable performance in complex, dynamic wireless environments, while keeping hardware costs and power consumption low.

The spatial flexibility offered by MAs has been effectively leveraged in several critical applications, including multi-user precoding \cite{MA_precode1, MA_precode2}, integrated beamforming for ISAC \cite{MA_ISAC2, MA_ISAC_MISO}, and efficient null steering for interference mitigation \cite{MA_null}. Furthermore, MAs have also played a pivotal role in enhancing physical layer security \cite{MA_PLS1, MA_PLS2_RIS2, MA_PLS3_MIMO} and in the deployment of RIS \cite{MA_PLS2_RIS2}, significantly broadening their potential for future wireless systems.
Meanwhile, Fluid Antenna Systems (FASs), which share a similar philosophy, focus on optimizing antenna positions within compact spatial regions to dynamically improve channel characteristics \cite{FA1, FA2, FA3}.
MAs and FASs enhance communication and sensing performance by dynamically adjusting antenna positions, optimizing channel conditions, and improving reliability, coverage, and throughput, all without the need for additional antennas or RF components. Their real-time adaptability supports applications such as autonomous driving, smart cities, and industrial automation, while working synergistically with RISs and mmWave technologies to reduce latency and improve system performance. These advantages make MAs and FASs particularly promising for next-generation wireless networks, especially in the context of 6G, where they align with sustainability goals by promoting energy efficiency and supporting green technologies. The integration of MAs and FASs, when combined with advanced algorithmic strategies, has revolutionized wireless communication by offering physical hardware adaptability, leading to substantial improvements in communication architectures such as Single-Input Single-Output (SISO) \cite{MA1_SISO}, Multiple-Input Single-Output (MISO) \cite{MA_ISAC_MISO, MA_MISO,MA5_MISO}, and Multiple-Input Multiple-Output (MIMO) \cite{MA_PLS3_MIMO,MA_MIMO4,MA3_MIMO}, thereby maximizing system performance and efficiency.

Before the advent of MAs and FASs, significant research in signal processing focused on optimizing antenna positions to enhance communication performance. 
Two primary techniques—Antenna Selection and Array Synthesis—laid the groundwork for spatial optimization. Antenna Selection involved selecting the optimal subset of antennas from a dense array to maximize metrics like ergodic capacity\cite{AS_1,AS_2}. This technique, although effective, required pre-deploying a large number of antennas and performing exhaustive Channel State Information (CSI) estimation for each element, leading to limitations in terms of resource efficiency.
 Array Synthesis aimed to tailor beam patterns to meet design criteria such as sidelobe suppression and enhanced main-lobe directivity through precise optimization of antenna positions, element counts, and beamforming coefficients\cite{ASyn1,ASyn2}. Although different in their approach and objectives, these early methods highlighted the significance of spatial optimization in communication systems.



In MIMO systems, MAs optimize performance not only by adjusting the positions of individual antennas at the single-element level but also by moving each antenna within the array at the array level.
The emergence of flexible substrates has enabled the deformation of array structures, leading to the development of flexible antenna arrays that allow for the movement of array antennas \cite{FS_FAA_1,MA_arraycylind,FS_FAA_2_c,FS_FAA_3,FS_FAA_4}.
 The evolution of MAs from single-element to array-level systems represents a significant advancement in wireless communication technology. This progression enhances the flexibility and performance of communication systems, effectively addressing challenges such as signal fading, interference, and capacity limitations.
  In single-element MAs, the dynamic adjustment of individual antenna positions optimizes signal reception and transmission, reducing the impact of channel impairments. However, transitioning to array-level MAs, which involve multiple antenna elements, further improves spatial diversity and multiplexing, leading to substantial gains in overall system performance.

 In the optimization of array-level MAs, significant advancements have already been made in the study of array antennas. \cite{MA_array6D1,MA_array6D2} proposed a 6-dimensional (6D) MA system that simultaneously accounts for both the position and rotation of the antenna array, thereby further enhancing system performance. Additionally, \cite{MA_arraylevel} discusses various general implementation architectures for array-level MAs in detail, underscoring their immense potential to revolutionize the field of wireless communication.
 Particularly, flexible cylindrical arrays, or dynamically adjustable cylindrical arrays, have been shown to offer advantages over traditional planar arrays. \cite{FS_FAA_2_c} proposed a flexible broadband dual-polarized low-profile conformal phased array antenna that achieves cylindrical bending through a modular architecture. Experimental results demonstrated the antenna's exceptional performance in wide-angle beam scanning and grating-lobe-free scanning, showcasing its potential for advanced communication systems. Similarly, \cite{MA_arraycylind} introduced a cylindrical conformal array antenna with a uniform cylindrical arrangement. This study explores the multifunctionality of linear antenna arrays when deployed on cylindrical surfaces, evaluating their radiation performance in terms of sidelobe levels, mutual coupling, and other key parameters.
 
In wireless communication applications, \cite{FFA} conducted a performance study on flexible cylindrical arrays, demonstrating the effectiveness of dynamically adjusting the flexible bending angle in enhancing communication system performance. 
In practical communication scenarios, dynamically adjustable cylindrical arrays offer significant advantages over traditional planar arrays. They provide seamless omnidirectional coverage, demonstrating the effectiveness of dynamic adjustment in enhancing communication system performance. This not only optimizes system performance but also simplifies physical design, positioning cylindrical arrays as a highly promising solution for modern wireless communication systems.




Although MAs have found numerous applications in wireless communication, there are few optimization schemes that focus on deploying MAs in cylindrical arrays. Moreover, research on cylindrical arrays has primarily concentrated on flexible substrates, with limited exploration of optimizing antenna performance in this new deployment context.
The use of cylindrical arrays for omnidirectional communication offers broader coverage compared to traditional planar arrays, providing distinct advantages for practical deployment in wireless communication systems.
In this paper, we explore the integration of an optimization scheme for MAs based on a Flexible Cylindrical Arrays (FCLA), which consists of multiple layers of Flexible Circle Arrays (FCA), aimed at improving communication rates and overall system performance in a Multi-User MISO (MU-MISO) downlink system.
 The system we proposed uses MAs based on Fractional Programming (FP) algorithms to optimize the multi-user sum rate of the system.
  We conduct extensive simulations to validate the effectiveness of the proposed solution, which demonstrates that the optimization scheme with movable antennas significantly outperforms traditional fixed antenna optimization, achieving a performance gain of up to 31\% in general scenarios. Below, we outline the specific novelties and contributions of our work:
\begin{itemize} 
	\item We propose a novel FCLA, a dynamic array system where antennas at the same height can revolve along a circular track. This FCLA enables both vertical and horizontal movement of the array elements, offering a significant advantage over traditional planar arrays. Specifically, in the horizontal dimension, antennas at each layer of the FCA can move along the track to dynamically adjust their revolving angles. In the vertical dimension, each FCA layer functions as a whole, allowing for dynamic adjustment of the layer's height. By leveraging these dynamic movements, the FCLA adapts in real-time to optimize wireless channels, enhancing system coverage and performance. 
	 
	\item We propose an optimization scheme using MAs under the FCLA architecture, based on the FP algorithm.
	We optimize the multi-user sum rate by adjusting the antenna positions to affect the wireless channel. The optimization problem is addressed using FP algorithms, which enables the alternating optimization of beamforming, revolving angles, and heights within a unified framework.
	This approach effectively provides a viable option for MA optimization, as it not only improves the system's overall performance but also allows for dynamic adjustment to changing network conditions.

	\item We propose a novel Constrained Grid Search-Based Adaptive Moment Estimation Algorithm (CGS-Adam) for antenna position optimization within the FP framework. We derive closed-form solutions for the gradient of the revolving angles and heights in the antenna position optimization process. 
	Based on these closed-form solutions, the proposed CGS-Adam algorithm integrates a momentum-based gradient strategy, adaptive step sizes, and grid search techniques to efficiently determine optimal antenna positions. 
	The algorithm ensures that all constraints, including the minimal inter-element spacing constraint (typically set to half the wavelength to prevent mutual coupling effects), are strictly adhered to during the optimization process.
	Specifically, during the optimization of antenna positions, alternating optimization is applied to both the horizontal and vertical dimensions. In each FCA, we optimize the revolving angle of the antennas within that FCA. By treating each FCA as a whole, we then optimize their heights. This alternating optimization strategy is applied to both the revolving angles and heights, ensuring that these two parameters are jointly optimized to enhance the overall system performance.
	The proposed CGS-Adam algorithm accelerates the convergence of antenna positions, enhancing the optimization process and improving efficiency.

\end{itemize}

The remainder of this paper is organized as follows. Section II develops a model for a base station (BS) with a cylindrical array of antennas, including the multi-user signal model and the multipath channel model. 
Section III provides the theoretical derivation of MAs optimization based on FP and proposes a specific optimization algorithm for its implementation.
 In Section IV, a comprehensive comparative analysis is conducted through extensive simulations to demonstrate the performance improvement achieved by applying MAs in the cylindrical array antenna distribution. Finally, Section V concludes the paper and outlines potential directions for future research.


{\emph {Notations}}:
${\left(  \cdot  \right)}^{\mathrm{T}}$, ${\left(  \cdot  \right)}^{\mathrm{H}}$, and $\left(\cdot\right)^{-1}$ denote   transpose, conjugate transpose, and inverse, respectively.
$|\cdot|$ denotes the modulus (or absolute value) of a matrix (or scalar).
\( \left[\mathbf{A}\right]_{i,j} \) denotes the \((i,j)\) element of matrix \( \mathbf{A} \).
$\mathbf{A}^\star$ denotes the optimal solution of $\mathbf{A}$.
$\Vert\mathbf{A}\Vert_F$ denotes the Frobenius norm of matrix $\mathbf{A}$. 
$\mathbb{E}\{\cdot\}$ denotes the expectation.  
$\Re\{\cdot\}$ denotes the real part of a complex number or matrix.
   

\section{System Model}
 
In this section, we will model the communication scenario of the BS, where the antennas are arranged in a cylindrical structure consisting of multiple circular layers with uniformly distributed elements.
To better describe the cylindrical distribution of antennas, a cylindrical coordinate system is established, with the center of the antenna distribution on the first layer serving as the origin. In this coordinate system, $r$ represents the radius, $\psi$ denotes the azimuthal angle, and $z$ represents the height.
  
\begin{figure*}[t] 
	\centering
	\includegraphics[width=0.8\linewidth]{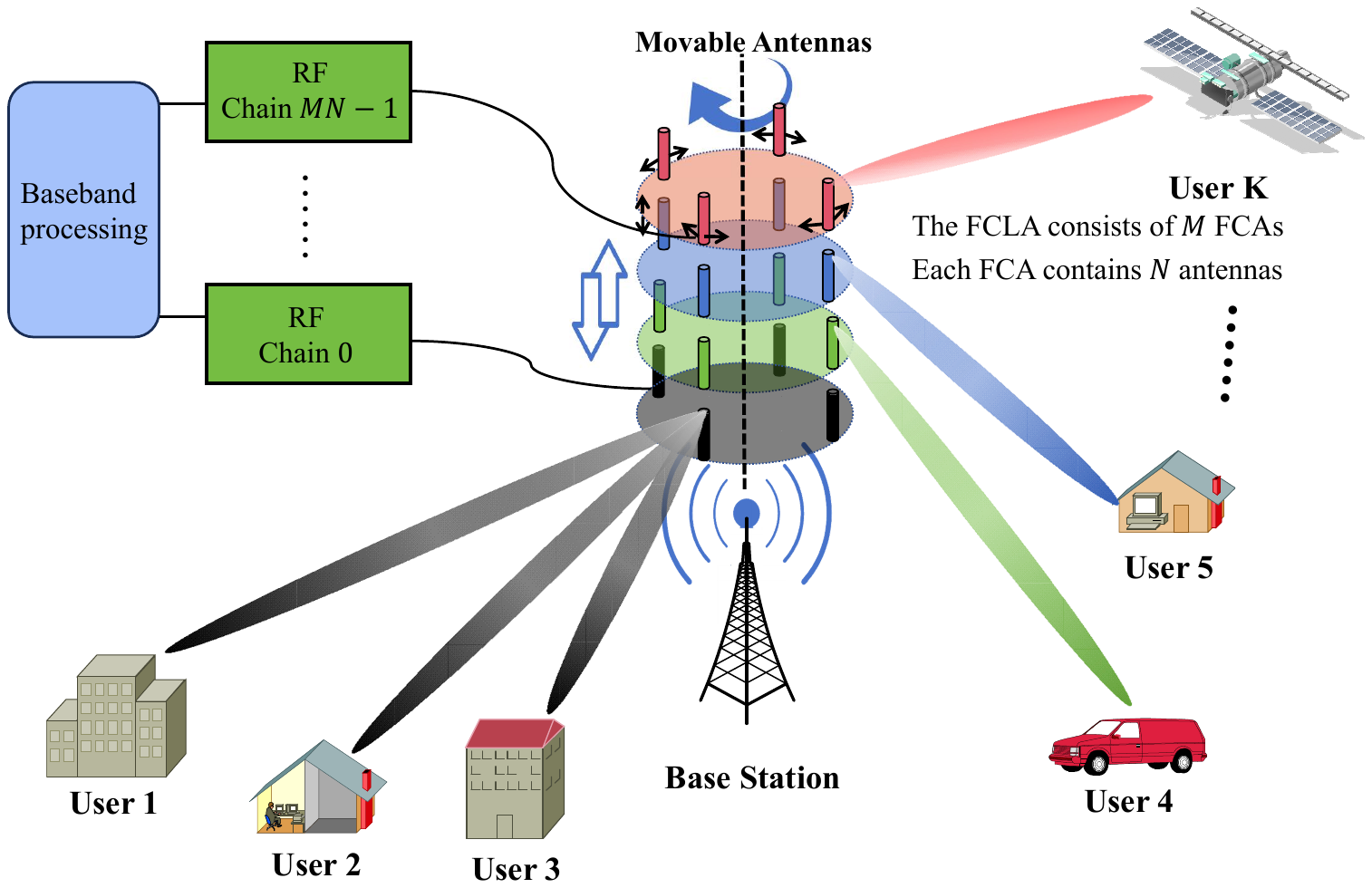} 
	\caption{Multiuser Communication System with Movable Antennas}
	\label{fig1}
\end{figure*}

We consider a MU-MISO downlink system, as illustrated in Fig. \ref{fig1}, where the BS is equipped with $M$ FCAs, with each layer deployed $N$ antennas that could orbit along the circle. The radius $R$ of each circular layer remains constant. $K$ users are equipped with a single antenna, and the BS supports a single data stream for each user. The $K$ users are distributed in a $360^\circ$ region around the BS.

For convenience, we represent the antenna angles and heights as follows: the revolving angle of the antenna located at the $j$-th position on the $i$-th layer is denoted by $\psi_{i,j}$, which is the element in the $i$-th row and $j$-th column of $\bm{\psi} \in \mathbb{R}^{M \times N}$. The height of the circular layer where the antennas of the $i$-th layer are positioned is represented by $z_i$, which is the $i$-th element of $\mathbf{z} \in \mathbb{R}^{M \times 1}$.

In the downlink communication system, the received signal at the $k$-th user, $k\in\{1,\cdots,K\}$, can be expressed by
\begin{equation}\label{eq1}
	y_k=\mathbf{h}_k^{\mathrm{H}}\mathbf{F}\mathbf{s}+n_k,
\end{equation}
where $\mathbf{h}_k\in\mathbb{C}^{MN\times 1}$ is the $k$-th user's channel, $\mathbf{F}\triangleq[\mathbf{f}_1,\cdots,\mathbf{f}_K]\in\mathbb{C}^{MN\times K}$ is the precoding matrix,
$\mathbf{s}\in\mathbb{C}^{K\times 1}$ represents the $K$ data streams for $K$ users, $n_k$ represents the Gaussian additive white noise following $\mathcal{CN}(0,\sigma_k^2)$.

By assuming the i.i.d. transmit data such that $\mathbb{E}\{s_k^*s_k\}=1$ and $\mathbb{E}\{s_i^*s_k\}=0$, $\forall k , k\neq i$,
the SINR of the $k$-user can be given by
\begin{equation}\label{eq2}
	{\rm SINR}_k=\frac{\vert\mathbf{h}_k^{\mathrm{H}}\mathbf{f}_k\vert^2}{\sum_{i,i\neq k}^{K}\vert\mathbf{h}_k^{\mathrm{H}}\mathbf{f}_i\vert^2+\sigma_k^2}.
\end{equation} 
By assuming $L$ paths for all user channels, the spatial channel under the far-field plane-wave assumption is given by, $\forall k$,
 \begin{equation}\label{eq3}
 	\mathbf{h}_k=\sqrt{\frac{1}{L}}\sum_{l=1}^{L}\beta_{k,l} \mathbf{a}(\vartheta_{k,l},\varphi_{k,l}),
 \end{equation}
where $\beta_{k,l}$, $k\in\{1,\cdots,K\}$, $l\in\{1,\cdots,L\}$, is the complex path gain of the $l$-th path of the $k$-user's channel,  $\vartheta_{k,l}$ and $\varphi_{k,l}$ correspond to the elevation and azimuth angles. The array-angle manifold $\mathbf{a}\in\mathbb{C}^{MN\times 1}$ is a column-stacked vector by $\mathbf{a}\triangleq[\mathbf{a}_1^\mathrm{T},\cdots,\mathbf{a}_{MN}^\mathrm{T}]^\mathrm{T}$, with $M$ and $N$ denoting the elevation and azimuth antenna counts, respectively, given by
\begin{equation}\label{eq4}
	\begin{aligned}
		\left[\mathbf{a}_m(\vartheta,\varphi)\right]_n
		=& e^{-j\frac{2\pi}{\lambda}( x_{m,n}\sin\vartheta\cos\varphi + y_{m,n}\sin\vartheta\sin\varphi  +z_m\cos\vartheta) }
		\\=&e^{-j\frac{2\pi}{\lambda}( R\cos\psi_{m,n}\sin\vartheta\cos\varphi + R\sin\psi_{m,n}\sin\vartheta\sin\varphi +z_m\cos\vartheta) } \\
		=& e^{-j\frac{2\pi}{\lambda}( \phi^x R \cos\psi_{m,n}+  \phi^y R \sin\psi_{m,n}+z_m \theta) } ,
	\end{aligned}
\end{equation}
where the second equation holds due to the antenna position mapping in the circle orbit, i.e., $x_n=R\cos\psi_n$ and $y_n=R\sin\psi_n$. Moreover,
 $\phi^x\triangleq \sin\vartheta\cos\varphi $, $\phi^y\triangleq \sin\vartheta\sin\varphi $, and $\theta\triangleq \cos\vartheta$ are defined as the virtual azimuth and elevation angles for clarity. 
 
In FCLAs, the flexible DoF for the revolving angles $\bm{\psi}\triangleq \{\psi_{m,n}\}_{m=1,n=1}^{MN}$ and heights $\bm{z}\triangleq \{z_{m}\}_{m=1}^{M}$ both impact the channel $\mathbf{h}_k, \forall m$, the SINR in Eqn. (\ref{eq2}) depends on both $\mathbf{F}$, $\bm{\psi}$ and $\bm{z}$. Meanwhile, considering both the power constraints and the position constraints to avoid the coupling effects of multiple antennas in the transmission area, the sum-rate optimization problem in FCLAs is formulated as follows:
\begin{subequations}\label{p5}
	\begin{equation}\label{p5_aa}
		\begin{aligned}
	&\underset{\mathbf{F},\bm{\psi},\bm{z}}{\rm arg\ max} \	\sum_{k=1}^{K}\log(1+\text{SINR}_k(\mathbf{F},\bm{\psi},\bm{z})) \\ 
		\end{aligned}
	\end{equation} 
	\begin{equation}\label{p5_bb}
		\begin{aligned}
	&{\rm s.t.}\ \Vert\mathbf{F}\Vert_F^2 \leq P,\\		
		\end{aligned}
	\end{equation}
	\begin{equation}\label{p6}
		\begin{aligned}
			{\rm s.t.} \ \Vert\mathbf{t_s}-\mathbf{t_{s'}}\Vert_F\geq \frac{\lambda_0}{2}, s\neq s',1 \leq s, s' \leq MN.
		\end{aligned}
	\end{equation}
\end{subequations}
where $P$ represents the total transmit power, $\mathbf{T} = [\mathbf{t}_1, \cdots, \mathbf{t}_{MN}]$ represents the position parameters of the antenna elements, $\lambda_0$ represents the wavelength of the antenna. 





\section{FP-Based Optimization of SINR}

In this section, we propose an FP algorithm with antenna position optimization to address the sum-rate maximization problem.
First, the fractional objective in problem (\ref{p5_aa}) is simplified into a more tractable equivalent form with respect to the optimization variables. Then, by employing the FP framework, the beamforming matrix $\mathbf{F}$ and the positions of the movable antennas, $\bm{\psi}$ and $\bm{z}$, are alternately updated while keeping the other variables fixed. This iterative optimization approach results in improved sum-rate performance.


To maximize the sum-rate, it is acknowledged that the inclusion of fractional terms in problem (\ref{p5_aa}) significantly complicates direct optimization, making it challenging to solve efficiently. To address these difficulties, fractional programming is employed to simplify the equation, effectively transforming the problem into a more tractable equivalent form \cite{FP1}. The position parameters \( \mathbf{T} \) are determined by \( \bm{\psi} \) and \( \bm{z} \). These angles must satisfy the angular spacing constraint to minimize mutual coupling among antennas, while also adhering to the inter-layer spacing constraint. 
Therefore, The problem (\ref{p5_aa}) is equivalent to
\begin{subequations}
\begin{equation}\label{8a}
	\begin{aligned}
		\underset{\mathbf{F}, \bm{\psi}, \bm{z}, \bm{\epsilon}, \bm{\mu}}{\rm arg\ max} \quad \mathcal{L} & = \sum_{k=1}^{K} \log(1 + \epsilon_k) - \sum_{k=1}^{K} \epsilon_k \\
		& + \sum_{k=1}^{K} (1 + \epsilon_k) [2\Re\{\mu_k^* a_k\} - |\mu_k|^2 b_k]
	\end{aligned}
\end{equation}
\begin{equation}\label{p7}
	\begin{aligned}
		 \text{s.t.} \quad &|\psi_{m,i} - \psi_{m,j}| \geq \psi_{\rm min}, |z_{i'} - z_{j'}| \geq z_{\min},\\
		 &  \forall m \in \{1, \cdots, M\}, \forall i,j \in \{1, \cdots, N\},\\
		 &  \forall i', j' \in \{1, \cdots, M\}, i \neq j, i' \neq j',\\
		 & \epsilon_k > 0, \; \mu_k \in \mathbb{C}, \; k \in \mathcal{N}, \text{Tr}(\mathbf{F} \mathbf{F}^\mathrm{H}) \leq P,
	\end{aligned}
\end{equation}
\end{subequations}
where $a_k = \mathbf{f}_k^{\mathrm{H}} \mathbf{h}_k$, $b_k = \sigma_k^2 + \sum_{i=1}^{K} |\mathbf{f}_i^{\mathrm{H}} \mathbf{h}_k|^2$, $\psi_{\rm min} = 2\arcsin\frac{\lambda_0}{4R}$, and $z_{\rm min} = \frac{\lambda_0}{2}$.
The optimal values of $\epsilon_k$ and $\mu_k$ are given by 
\begin{equation} \label{eq9}
	\epsilon_k^{\star} = \frac{|a_k|^2}{\sum_{k' \neq k} |\mathbf{f}_{k'}^{\mathrm{H}} \mathbf{h}_k|^2 + \sigma_k^2} = \gamma_k,
\end{equation}
\begin{equation}\label{eq10}
	 \mu_k^{\star} = a_k b_k^{-1}.
\end{equation}

Problem (\ref{8a}) involves the optimization of five variables. We adopt an alternating iterative optimization approach, where each variable is optimized while keeping the others fixed. Since the optimal solutions for \( \bm{\epsilon} \) and \( \bm{\mu} \) have already been given by \( \mathbf{F} \), \( \bm{\psi} \), and \( \bm{z} \), we focus on optimizing \( \mathbf{F} \), \( \bm{\psi} \), and \( \bm{z} \).

\subsection{Optimization of the Beamforming Matrix $\mathbf{F}$}

To optimize $\mathbf{F}$ while keeping the other variables fixed, we start from problem (\ref{8a}) and derive the following optimization problem
\begin{equation}\label{eq11}
	\begin{aligned}
		\underset{\text{Tr}(\mathbf{F} \mathbf{F}^\mathrm{H}) \leq P}{\rm arg\ min} q(\mathbf{F}) & = \sum_{k=1}^{K} (1 + \epsilon_k) \left( |\mu_k|^2 b_k - 2 \Re \left\{ \mu_k^* a_k \right\} \right) \\
		 					& = \sum_{k=1}^{K} (1 + \epsilon_k) \left( |\mu_k|^2 \sigma_k^2 \right) \\
			 					& + \sum_{i=1}^{K} \mathbf{f}_i^{\mathrm{H}} \left[ \sum_{k=1}^{K} |\mu_k|^2 (1 + \epsilon_k) \mathbf{h}_k \mathbf{h}_k^{\mathrm{H}} \right] \mathbf{f}_i \\
		 					& - 2 \Re \left\{ \sum_{k=1}^{K} (1 + \epsilon_k) \mu_k^* \mathbf{f}_k^{\mathrm{H}} \mathbf{h}_k \right\}.					
	\end{aligned}
\end{equation}

Rewriting problem (\refeq{eq11}) in matrix form, we obtain:
\begin{subequations}
	\begin{equation}
		q(\mathbf{F}) = \operatorname{Tr}(\mathbf{M}) + \operatorname{Tr}(\mathbf{F}^\mathrm{H} \mathbf{C} \mathbf{F}) - 2\Re\left\{ \operatorname{Tr}(\mathbf{F}^\mathrm{H} \mathbf{D}) \right\},
	\end{equation}
	where
	\begin{equation}
		\mathbf{C} = \sum_{k=1}^{K} (1 + \epsilon_k) |\mu_k|^2 \mathbf{h}_k \mathbf{h}_k^{\mathrm{H}},
	\end{equation}
	\begin{equation}
		\mathbf{D} = \left[ (1 + \epsilon_1) \mu_1^* \mathbf{h}_1, \cdots, (1 + \epsilon_K) \mu_K^* \mathbf{h}_K \right],
	\end{equation}
	\begin{equation}
		\mathbf{M} = \operatorname{diag} \{ (1 + \epsilon_1) |\mu_1|^2 \sigma_1^2, \cdots, (1 + \epsilon_K) |\mu_K|^2 \sigma_K^2 \}.
	\end{equation}
\end{subequations}

Thus, the optimal solution for $\mathbf{F}^\star$ can be expressed as:
\begin{equation}
	\min_{\operatorname{Tr}(\mathbf{F} \mathbf{F}^\mathrm{H}) \leq P} \left( \operatorname{Tr}(\mathbf{F}^\mathrm{H} \mathbf{C} \mathbf{F}) - 2 \Re\{\operatorname{Tr}(\mathbf{F}^\mathrm{H} \mathbf{D})\} + \operatorname{Tr}(\mathbf{M}) \right),
\end{equation}
this is a constrained convex optimization problem, where $\mathbf{M}$ is a constant independent of $\mathbf{F}$, employing the Lagrange multiplier method. The Lagrangian function is constructed to incorporate the constraints into the objective function. We can obtain:
\begin{equation}
	g(\mathbf{F}) = \text{Tr}(\mathbf{F}^\mathrm{H} \mathbf{C} \mathbf{F}) - 2 \Re \{ \text{Tr}(\mathbf{F}^\mathrm{H} \mathbf{D}) \} + \text{Tr}(\mathbf{M}) + \lambda (\text{Tr}(\mathbf{F}^\mathrm{H} \mathbf{F}) - P),
\end{equation}
where $\lambda$ represents the Lagrange multiplier. Take the derivative of $g(\mathbf{F})$ to obtain the optimal solution that satisfies the constraint conditions, as shown in the following expression:
\begin{equation}\label{eq15}
	\mathbf{F}^\star = (\mathbf{C} + \lambda \mathbf{I})^{-1} \mathbf{D},
\end{equation}
where the constraint conditions Eqn. (\refeq{p5_bb}) must be satisfied. If the power constraint $\text{Tr}(\mathbf{F} \mathbf{F}^\mathrm{H}) \leq P$ holds, then $\lambda = 0$. However, if the constraint is not met, specifically when $\text{Tr}(\mathbf{F} \mathbf{F}^\mathrm{H}) > P$, $\lambda$ must be adjusted to fulfill the power constraint condition.Inserting Eqn. (\refeq{eq15}) into Eqn. (\refeq{p5_bb}) yields the following expression:
\begin{equation}\label{eq16}
	\begin{aligned}
	\text{Tr}(\mathbf{F} \mathbf{F}^\mathrm{H}) &= \text{Tr}((\mathbf{C} + \lambda \mathbf{I})^{-1} \mathbf{D} \mathbf{D}^\mathrm{H} (\mathbf{C} + \lambda \mathbf{I})^{-1}) \\
												&= \text{Tr}(\mathbf{D}^\mathrm{H} (\mathbf{C} + \lambda \mathbf{I})^{-2} \mathbf{D})  \\
												&= P.
	\end{aligned}
\end{equation}

Perform the eigenvalue decomposition of $\mathbf{C}$ as $\mathbf{C} = \mathbf{P}^\mathrm{H} \mathbf{\Lambda} \mathbf{P}$, and then substitute into Eqn. (\refeq{eq16}) to obtain the following expression:
\begin{equation}\label{eq17}
	\sum_{n=1}^{MN} \frac{[\mathbf{P} \mathbf{D} \mathbf{D}^\mathrm{H} \mathbf{P}^\mathrm{H}]_{n,n}}{([\mathbf{\Lambda}]_{n,n} + \lambda)^2} = P,
\end{equation}
where \( [\mathbf{\Lambda}]_{n,n} \geq 0 \) for \( n \in \mathcal{N} \), the left-hand side of Eqn. (\refeq{eq17}) behaves as a monotonic function with respect to \( \lambda \geq 0 \). Consequently, \( \lambda \) can be determined by solving Eqn. (\ref{eq17}) using a bisection search method, thereby ensuring that the constraint conditions are satisfied.

\subsection{Optimization of Antenna Positions \texorpdfstring{$\bm{\psi}$}{$\psi$} and $\mathbf{z}$}

After obtaining the optimal solution for \( \mathbf{F^\star} \), the matrix \( \mathbf{H} \triangleq [\mathbf{h}_1, \cdots, \mathbf{h}_K] \in \mathbb{C}^{MN \times K} \) also influences the SINR. By adjusting the antenna positions to change the values of \( \bm{\psi} \) and \( \mathbf{z} \) for the antennas, \( \mathbf{h}_k \) can be altered, thereby enhancing the optimization performance.

Keeping  $\boldsymbol{\epsilon}$, $\boldsymbol{\mu}$ and $\mathbf{F}$  constant, we obtain the following from problem (\ref{8a}):
\begin{equation}\label{eq18}
	\begin{aligned}
		\underset{\text{Eqn.} (\refeq{p7}) }{\rm arg\ max} f(\bm{\psi},\mathbf{z}) = &\sum_{k=1}^{K} \left[ 2 \Re \{ (1 + \epsilon_k) \mu_k^* \sum_{n=1}^{MN} f_{k,n}^* h_{k,n} \} \right. \\
		&- (1 + \epsilon_k) |\mu_k|^2 \sigma_k^2 \\
		&- (1 + \epsilon_k) |\mu_k|^2 \mathbf{h}_k^\mathrm{H} \sum_{i=1}^{K} \mathbf{f}_i \mathbf{f}_i^\mathrm{H} \mathbf{h}_k \bigg] \\
		{\rm s.t.} \ & \;   \mathrm{Eqn.} (\refeq{p7})  \;.
	\end{aligned}
\end{equation}

Problem (\ref{eq18}) is a function of the sum over all $MN$ antennas. We can rewrite it in the form that describes the optimization for a single antenna, as follows:
\begin{equation}\label{eq19}
	 f(\bm{\psi},\mathbf{z}) = \sum_{s=1}^{MN} f_s(\psi_{m,n},z_m),
\end{equation}
where $s = (m-1)N + n$, $0 \leq s \leq MN$, \( 0 \leq m \leq M \), \( 0 \leq n \leq N \), with \( m \in \mathcal{N} \) representing the \( m \)-th layer and \( n \in \mathcal{N} \) denoting the \( n \)-th antenna within a given layer.

The expression in terms of each individual antenna is shown as follows:
\begin{equation}\label{eq20}
	\begin{aligned}
		& f_s(\psi_{m,n},z_m) = \sum_{k=1}^{K} \left( 2 \Re \{ h_{k,s}^*(\psi_{m,n},z_m) c_{k,s} \} \right. \\
		& \quad - d_{k,s} |h_{k,s}(\psi_{m,n},z_m)|^2 - \frac{(1 + \epsilon_k) |\mu_k|^2 \sigma_k^2}{MN}).
	\end{aligned}
\end{equation}

The expression for $c_{k,s}$, $d_{k,s}$ is given as follows:
\begin{equation}
	c_{k,s} = (1 + \epsilon_k) ( \mu_k f_{k,s} - \frac{1}{2} |\mu_k|^2 \sum_{s' \neq s}^{MN} F_{s,s'} h_{k,s'}(\psi_{s'},z_{m'})),
\end{equation}
\begin{equation}
	 d_{k,s} = (1 + \epsilon_k) |\mu_k|^2 F_{s,s},
\end{equation}
where \( F_{i,j} \) is the \((i,j)\)-th element of the matrix \( \sum_{k=1}^{K} \mathbf{f}_k \mathbf{f}_k^{\mathrm{H}} \). For the optimization problem in problem (\refeq{eq18}), we apply the gradient ascent method to optimize it. Expressing $ h_{k,s}(\psi_{m,n}, z_{m}) $ in explicit form as \( h_{k,s}(\psi_{m,n},z_{m}) =\sqrt{\frac{1}{L}} \sum_{l=1}^{L} \beta_{k,l} e^{-j \frac{2 \pi}{\lambda} ( \phi_{k,l}^x R \cos \psi_{m,n} + \phi_{k,l}^y R \sin \psi_{m,n} + z_m \theta_{k,l} )} \) and substituting it into Eqn. (\refeq{eq20}). We can obtain:

\begin{equation}\label{eq23}
		\renewcommand{\arraystretch}{1.5}
		\nabla_{(r, \psi, z)} f_s = \begin{bmatrix} 
			\frac{\partial f_s}{\partial r} \\ 
			\frac{1}{r} \frac{\partial f_s}{\partial \psi} \\ 
			\frac{\partial f_s}{\partial z} 
		\end{bmatrix}
		 = \begin{bmatrix} 
			0 \\ 
			\frac{1}{R} \frac{\partial f_s}{\partial \psi_{m,n}} \\ 
			\frac{\partial f_s}{\partial z_{m}} 
		\end{bmatrix},
\end{equation}
where 
\begin{equation}\label{eq24}
	\begin{aligned}
	\frac{1}{R}& \frac{\partial f_s}{\partial \psi_{m,n}} = \sqrt{\frac{1}{L}} \sum_{k=1}^K \sum_{l=1}^{L} \frac{\left|\beta_{k,l}^* c_{k,s}\right|}{\frac{\lambda}{4 \pi}} \Psi_{k,l}(\psi_{m,n}) \cdot\\
	& \sin \left( \frac{2 \pi}{\lambda} \mathbf{t}_s^\mathrm{T} \bm{\varXi}_{k,l}  + \angle (\beta_{k,l}^* c_{k,s}) \right) \\
	& +\frac{1}{L} \sum_{k=1}^K \sum_{l=1}^{L} \sum_{l' \neq l}^{L} 
	\frac{|\beta_{k,l} \beta_{k,l'}^*|}{\frac{\lambda}{2 \pi d_{k,s}}} (\Psi_{k,l}(\psi_{m,n}) - \Psi_{k,l'}(\psi_{m,n})) \cdot\\
	& \sin \left( \frac{2 \pi}{\lambda} \mathbf{t}_s^\mathrm{T} (\bm{\varXi}_{k,l} - \bm{\varXi}_{k,l'}) + \left( \angle \beta_{k,l'} - \angle \beta_{k,l} \right) \right),
	\end{aligned}
\end{equation}
\begin{equation}
	\begin{aligned}
	\frac{\partial f_s}{\partial z_{m}} &= \sqrt{\frac{1}{L}} \sum_{k=1}^K \sum_{l=1}^{L} \frac{|\beta_{k,l}^* c_{k,s}|}{-\frac{\lambda}{4 \pi}} \theta_{k,l} \sin(\frac{2 \pi}{\lambda}\mathbf{t}_s^\mathrm{T} \bm{\varXi}_{k,l} + \angle (\beta_{k,l}^* c_{k,s}))   \\
	&+ \frac{1}{L} \sum_{k=1}^K \sum_{l=1}^{L} \sum_{l' \neq l}^{L} 
	\frac{|\beta_{k,l} \beta_{k,l'}^*|}{\frac{\lambda}{2 \pi d_{k,s}}} (\theta_{k,l} - \theta_{k,l'}) \cdot\\
	& \sin \left( \frac{2 \pi}{\lambda} \mathbf{t}_s^\mathrm{T} (\bm{\varXi}_{k,l} - \bm{\varXi}_{k,l'}) + \left( \angle \beta_{k,l'} - \angle \beta_{k,l} \right) \right),
	\end{aligned}
\end{equation}
where $\Psi_{k,l}(\psi_{m,n}) = \phi_{k,l}^x \sin \psi_{m,n} - \phi_{k,l}^y \cos \psi_{m,n}$, $\bm{\varXi}_{k,l} =\begin{bmatrix} \phi_{k,l}^x & \phi_{k,l}^y & \theta_{k,l} \end{bmatrix}^\mathrm{T}$, representing the incident azimuth angle.
$\mathbf{t}_s =\begin{bmatrix} R \sin \psi_{m,n} &  R \cos \psi_{m,n} & z_m \end{bmatrix}^\mathrm{T}$ represents the antenna position. 

According to Eqn. (\refeq{eq19}), the gradient expression of $f(\bm{\psi}, \mathbf{z})$ is as follows:
\begin{equation} \label{eq26}
	\renewcommand{\arraystretch}{1.5}
	\nabla_{(r, \psi, z)} f = \begin{bmatrix} 
		0 \\ 
		\frac{1}{R} \frac{\partial f_s}{\partial \psi_{m,n}} \\ 
		\frac{\partial f_s}{\partial z_{m}} 
	\end{bmatrix}
	+\begin{bmatrix} 
		0 \\ 
		\frac{1}{R} \sum_{s' \neq s}^{MN}\frac{\partial f_{s'}}{\partial \psi_{m,n}}\\ 
		\sum_{s' \neq s}^{MN}\frac{\partial f_{s'}}{\partial z_m}
	\end{bmatrix},
\end{equation}
where the first term is given by Eqn. (\refeq{eq23}), and by substituting Eqn. (\refeq{eq20}), the specific expression for the second term can be obtained:
\begin{equation}
	\begin{aligned}
		& \frac{1}{R} \sum_{s' \neq s}^{MN}\frac{\partial f_{s'}}{\partial \psi_{m,n}}  =\frac{1}{R} \sum_{s' \neq s}^{MN} \sum_{k=1}^{K} ( 2 \Re \{ h_{k,s'}^*(\psi_{m',n'},z_{m'}) \frac{\partial c_{k,{s'}}}{\partial \psi_{m,n}} \} )  \\
		& =\frac{1}{L} \sum_{s' \neq s}^{MN} \sum_{k=1}^{K} \sum_{l=1}^{L} \sum_{l' = 1}^{L} \frac{2 \pi}{\lambda}(1 + \epsilon_k) |\mu_k|^2 |F_{s',s}\beta_{k,l}^*\beta_{k,l'}| \cdot \\
		& \Psi_{k,l'}(\psi_{m,n}) \sin(\frac{2 \pi}{\lambda} (\mathbf{t}_{s'}^\mathrm{T} \bm{\varXi}_{k,l} - \mathbf{t}_{s}^\mathrm{T} \bm{\varXi}_{k,l'})+  \angle (F_{s',s}\beta_{k,l}^*\beta_{k,l'})).
	\end{aligned}
\end{equation}

Similarly, we can obtain:
\begin{equation}
	\begin{aligned}
	& \sum_{s' \neq s}^{MN}\frac{\partial f_{s'}}{\partial z_m} = \sum_{s' \neq s}^{MN} \sum_{k=1}^{K} ( 2 \Re \{ h_{k,s'}^*(\psi_{m',n'},z_{m'}) \frac{\partial c_{k,{s'}}}{\partial z_m} \} )  \\
	& =\frac{1}{L} \sum_{s' \neq s}^{MN} \sum_{k=1}^{K} \sum_{l=1}^{L} \sum_{l' = 1}^{L} \frac{2 \pi}{\lambda}(1 + \epsilon_k) |\mu_k|^2 |F_{s',s}\beta_{k,l}^*\beta_{k,l'}| \cdot \\
	& \theta_{k,l'} \sin(\frac{2 \pi}{\lambda} (\mathbf{t}_{s'}^\mathrm{T} \bm{\varXi}_{k,l} - \mathbf{t}_{s}^\mathrm{T} \bm{\varXi}_{k,l'})+  \angle (F_{s',s}\beta_{k,l}^*\beta_{k,l'})).
	\end{aligned}
\end{equation}
%

Thus, the gradient expression of Eqn. (\refeq{eq19}) is obtained. Considering the current real-world scenario, the $z_m$ of antennas at each layer should be the same. Therefore, directly applying gradient ascent optimization to $f$ with respect to both $\psi_{mn}$ and $z_m$ simultaneously would not ensure consistent antenna heights across layers. To address this, we adopt an Alternating Optimization approach: by fixing $\mathbf{z}$, we perform single-variable optimization on $\bm{\psi}$, and by fixing $\bm{\psi}$, we perform single-variable optimization on $\mathbf{z}$.

Due to the limitations of the gradient ascent algorithm, it generally yields only a local optimum rather than a global optimum. Therefore, in the optimization process, we employ a Grid Search with Gradient Ascent approach. In the alternating iteration, taking the optimization of $\bm{\psi}$ with fixed $\mathbf{z}$ as an example, we initially sample the positions of \( V \) points on a circular ring and calculate the function values \( f_v \) at these points. Gradient ascent optimization is then initiated from the position with the maximum function value. This approach enables local optimization within the identified optimal region, allowing for a more precise solution and enhancing the overall optimization performance.

\begin{algorithm} 
	\caption{The Constrained Grid Search-Based Adam Algorithm for Optimizing $\bm{\psi}$} 
	\label{FP_grid_psi}
	\KwData {$\mathbf{H}$, $\mathbf{F}$, $\bm{\psi}$, $\mathbf{z}$, $\alpha$, $I_g$, $tol$, $\bm{\epsilon}$, $\bm{\mu}$, $M$, $N$, $K$, $L$, $R$}   
	\KwResult {Optimized position $\bm{\psi}^\star$ and channel $\mathbf{H}^\star$.}         			   							
	\BlankLine
	\Begin{ 
		$\textbf{Initialization:}$ $I$, $V$, $\beta_1$, $\beta_2$, $\eta$ 			\\               
		\Repeat{convergence or $I$ is reached}{
			\For{$m=1,\cdots,M$}{ 
				\For{$n=1,\cdots,N$}{\For{$\psi_{m,n}$ in $\Psi_{range}$}{
						\If{$\vert\psi_{m,n}-\psi_{m,pre}^\star \vert < \psi_{\rm min}$}{ continue }
						Compute $f(\psi_{m,n})$ using Eqn. (\refeq{eq19}\\
						Find $f_{\text{max}}(\psi_{m,n})$ via grid search \\		
						\Repeat{convergence or $I_g$ is reached}{Compute $\frac{1}{R} \frac{\partial f}{\partial \psi_{m,n}}$ based on Eqn. (\refeq{eq26}) \\
							Compute the Adam parameters $\hat{m}_t$, $\hat{v}_t$ based on Eqn. (\refeq{eq31}) \\
							Update $\psi_{m,n}$ based on Eqn. (\refeq{eq32})}
						\If{$\vert\psi_{m,n}-\psi_{m,pre}^\star \vert < \psi_{\rm min}$}{Retain the initial value}
				}}
			}  
			
		}
	}        \Return{$\mathbf{H^\star}$, $\bm{\psi}^\star$}
\end{algorithm}

In the gradient ascent algorithm, we employ the Adaptive Moment Estimation (Adam) method, which combines momentum-based gradient ascent with adaptive step sizes to achieve a more accurate optimal solution by dynamically adjusting the learning rate through estimates of the first and second moments. Adam integrates momentum and smoothed adaptive learning rate adjustments, which accelerates convergence and reduces the likelihood of getting trapped in local optima. The calculation of the momentum, which represents the first moment estimate of the gradient, is given as follows:
\begin{equation}
	m_t = \beta_1 m_{t-1} + (1 - \beta_1) g_t,
\end{equation}
where $t$ denotes the iteration count, i.e., the time step. $m_t$ denotes the first moment estimate of the gradient, i.e., the momentum term, at the current time step, and $\beta_1 \in [0, 1)$ is the exponential decay rate for the first moment, typically set to 0.9. $g_t$ represents the gradient value at the current time step, specifically expressed as $\frac{1}{R} \frac{\partial f}{\partial \psi_{m,n}}$ or $\frac{\partial f}{\partial z_m}$.

The cumulative update of the squared gradient, i.e., the second moment estimate of the gradient, can be expressed as:
\begin{equation}
	v_t = \beta_2 v_{t-1} + (1 - \beta_2) g_t^2,
\end{equation}
where $v_t$ represents the second moment estimate, and $\beta_2 \in [0, 1)$ is the exponential decay rate for the second moment, typically set to 0.999.

Since the values of the first and second moment estimates are relatively small in the initial stages, especially during the first few iterations, the computed first and second moments tend to be biased toward zero, leading to slow parameter updates. Therefore, bias correction is introduced to adjust $m_t$ and $v_t$, making them more accurate in the initial stages. The bias correction expression is as follows:
\begin{equation}\label{eq31}
	\hat{m}_t = \frac{m_t}{1 - \beta_1^t}, \quad \hat{v}_t = \frac{v_t}{1 - \beta_2^t},
\end{equation}
where $\hat{m}_t$ and $\hat{v}_t$ represent the bias-corrected first and second moments of the gradient, respectively. Since we aim to maximize the function, the final parameter update is given by
\begin{equation}\label{eq32}
	\psi_{m,n}^{(t)} = \psi_{m,n}^{(t-1)} + \frac{\alpha \hat{m}_t}{\sqrt{\hat{v}_t} + \eta}, z_{m}^{(t)} = z_{m}^{(t-1)} + \frac{\alpha \hat{m}_t}{\sqrt{\hat{v}_t} + \eta},
\end{equation}
where $\alpha$ is the step size, i.e., the learning rate, and $\eta$ is a small positive constant to prevent division by zero, typically set to $10^{-8}$.

We apply the CGS-Adam approach to solve the antenna position optimization problem, where the position of each antenna is optimized sequentially.
When optimizing the position of a specific antenna, the positions of all other antennas remain fixed. This process ultimately yields the optimal positions for all antennas. As analyzed from Eqn. (\refeq{eq26}), $f$ is related to the initial positions of the antennas. Therefore, after obtaining the optimal positions for all antennas, multiple iterations are required until convergence to achieve the optimal solution, with a maximum number of iterations $I$.

During the optimization process, the optimized antenna positions should satisfy the constraint condition, i.e., Eqn. (\refeq{p7}), with respect to the previously optimized antenna positions. Specifically, in the grid search, if the current optimal solution does not satisfy the constraints, it is discarded, and the next best solution is chosen. 
In the gradient ascent process, if the optimal solution does not meet the constraints, the current solution is discarded, and the position is adjusted to maintain the minimum distance that satisfies the constraints.
The algorithm diagram for optimizing $\bm{\psi}$ is shown in Algorithm \ref{FP_grid_psi}. The maximum number of iterations for gradient ascent is $I_g$. The convergence threshold is denoted as $tol$. 
The optimization algorithm for $\mathbf{z}$ follows the same procedure as the \( \bm{\psi} \) algorithm. Due to the constraints, during the optimization of $\mathbf{z}$, it is only necessary to iterate over the number of layers \( M \), without the need to iterate over the number of antennas \( N \). The algorithm flowchart is presented in Algorithm \ref{FP_grid_z}.

%

\begin{algorithm} 
	\caption{The Constrained Grid Search-Based Adam Algorithm for Optimizing $\mathbf{z}$} 
	\label{FP_grid_z}
	\KwData {$\mathbf{H}$, $\mathbf{F}$, $\mathbf{z}$, $\bm{\psi}$, $\alpha$, $I_g$, $tol$, $\bm{\epsilon}$, $\bm{\mu}$, $M$, $N$, $K$, $L$, $R$}   
	\KwResult {Optimized position $\bm{z}^\star$ and channel $\mathbf{H}^\star$.}         			   							
	\BlankLine
	\Begin{ 
		$\textbf{Initialization:}$ $I$, $V$, $\beta_1$, $\beta_2$, $\eta$ 			\\               
		\Repeat{convergence or $I$ is reached}{
			\For{$m=1,\cdots,M$}{ 
				\For{$z_{m}$ in $z_{pre}^\star + z_{range}$}{
					Compute $f(z_{m})$ using Eqn. (\refeq{eq19})\\
					Find $f_{\text{max}}(z_{m})$ via grid search \\		
					\Repeat{convergence or $I_g$ is reached}{Compute $\frac{\partial f}{\partial z_m}$ based on Eqn. (\refeq{eq26}) \\
						Compute the Adam parameters $\hat{m}_t$, $\hat{v}_t$ based on Eqn. (\refeq{eq31}) \\
						Update $z_{m}$ based on Eqn. (\refeq{eq32})}
					\If{$\vert z_{m}-z_{pre}^\star \vert < z_{\rm min}$}{Retain the initial value}
				}
			}  
			
		}
	}        \Return{$\mathbf{H^\star}$, $\bm{z}^\star$}
\end{algorithm}

\subsection{Alternating Iterative Optimization of the Beamforming Matrix and Antenna Positions}

Once the optimal antenna positions are obtained, the optimal channel matrix $ \mathbf{H} $ is determined. The iterative algorithm based on FP is then applied, with the algorithm flowchart shown in Algorithm \ref{FP_Based Alg}. The beamforming matrix $\mathbf{F}$ and antenna positions, $\bm{\psi}$ and $\mathbf{z}$, are alternately optimized through iterative updates. The algorithm iterates multiple times until the sum rate converges to a steady-state solution or reaches the maximum number of iterations \( I_{fp} \).

\begin{algorithm} 
	\caption{FP-Based Algorithm for Optimizing Movable Antennas} 
	\label{FP_Based Alg}
	\KwData {$\bm{\psi}_0$, $\mathbf{z}_0$, $\mathbf{F}_0$, $M$, $N$, $K$, $L$, $R$, $P$}   
	\KwResult {Optimized the Beamforming Matirx $\mathbf{F}^\star$ and the Channel $\mathbf{H}^\star$.}        	
	\BlankLine
	\Begin{ 
		$\textbf{Initialization:}$ $I_{fp}$, the current iteration $i = 1$ 			\\               
		\Repeat{convergence or $I_{\text{fp}}$ is reached}{
			Compute $\bm{\epsilon}_i$ and $\bm{\mu}_i$ based on Eqn. (\refeq{eq9}) and Eqn. (\refeq{eq10}) \\
			Optimise $\mathbf{F}_{i-1}$ using $\bm{\epsilon}_i$ $\bm{\mu}_i$ $\bm{\psi}_{i-1}$ $\mathbf{z}_{i-1}$, and obtain the optimized $\mathbf{F}_i$ \\
			Optimise $\bm{\psi}_{i-1}$ and $\mathbf{z}_{i-1}$ based on Eqn. (\refeq{eq26}) using CGS-Adam, and alternately iterate to optimize \(\bm{\psi}\) and \(\mathbf{z}\), obtaining the optimized $\bm{\psi}_{i}$ and $\mathbf{z}_{i}$ \\
			Update $i = i + 1$
		}
	}        \Return{$\mathbf{F^\star}$, $\mathbf{H^\star}$}
\end{algorithm}

\section{Simulation Results} 

The system operates at a central frequency of $3$ GHz. The BS is equipped with $M=4$ layers, each containing $N=4$ antennas, serving $K=4$ users. Both the total transmit power and the noise power are standardized to $1$, and the number of propagation paths is \( L = 11 \) . The user and scatterer locations are uniformly distributed with azimuth angles $\varphi_{k,l}$ and elevation angles $\vartheta_{k,l}$ ranging within $[0,\pi]$ for all $k,l$. All users are assumed to have an identical number of channel paths and equal power.
The initial antenna positions are set such that each FCA contains \( N \) antennas uniformly distributed along a circle, with each FCA evenly spaced by one wavelength in the vertical dimension.
 Monte Carlo simulations are conducted for each experiment.
 The key methods included in this evaluation are:
 \begin{itemize} 
 	\item \textbf{Fixed Antenna Position Based on FP}: The antenna positions are kept fixed. The FP approach is applied to optimize only the beamforming matrix \( \mathbf{F} \), iterating until the sum-rate converges or the maximum number of iterations is reached.

 	
 	\item \textbf{Movable Antenna Position Based on CGS-Adam}: The antenna positions are allowed to move, with \( N \) antennas in each layer capable of revolving at different angles, and the height of each layer is adjustable, incorporating our proposed approach as detailed in Algorithm \ref{FP_Based Alg}. The antenna position optimization uses CGS-Adam, that is, alternately optimizing the positions, denoted as \( \bm{\psi} \) and \( \mathbf{z} \), as detailed in Algorithm \ref{FP_grid_psi} and Algorithm \ref{FP_grid_z}. The positions of the antennas and the beamforming matrix \( \mathbf{F} \) are iteratively optimized until the sum-rate converges or the maximum number of iterations is reached.

 	\item \textbf{Movable Antenna Position Based on Grid Search}: Similarly, the position of the antenna can be adjusted using Algorithm \ref{FP_Based Alg}. However, the optimization of antenna positions, including both the revolving angles and heights, is performed using a grid search method, without incorporating gradient ascent.
 	
	\item \textbf{Movable Antenna Position in Different Dimensions}: The position of the antenna can be adjusted, but the optimization process exclusively considers either the horizontal dimension (antenna revolving angle) or the vertical dimension (antenna height) in order to evaluate the performance improvement.
 \end{itemize} 

\begin{figure}
	\centering
	\includegraphics[width=0.98\linewidth]{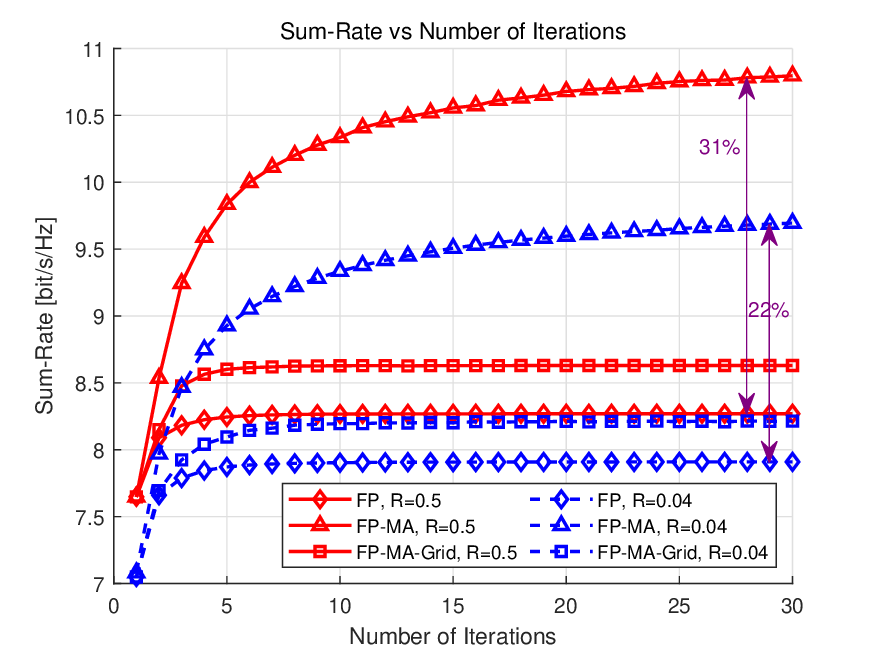}
	\caption{Sum-Rate vs. Iterations for Different Position Optimization Schemes and Horizontal Movable Ranges}
	\label{fig_iterR}
\end{figure}

The first simulation compares the performance of the sum-rate for fractional programming with fixed antenna positions and fractional programming with movable antenna positions as the number of iterations increases. The antenna movement is evaluated using two different schemes, as shown in Fig. \ref{fig_iterR}. 
The meanings of each curve are explained as follows:
\begin{itemize}
	\item \textbf{FP}: Represents the Fixed Antenna Position Based on the FP scheme without MAs.
	\item \textbf{FP-MA}: Represents the Movable Antenna Position Based on CGS-Adam scheme.
	\item \textbf{FP-MA-Grid}: Represents the Movable Antenna Position Based on grid search scheme without incorporating Adam optimization.
\end{itemize}
This comparison is performed when the antenna movement radius \( R = 0.5 \). A total of 30 iterations of \( I_{fp} \) are conducted. The first iteration records the initial values. As shown, our proposed approach significantly outperforms the fixed antenna FP algorithm. After the sum-rate converges, the performance improves by 31\% compared to the FP algorithm for fixed antenna position. This improvement is of great significance in the practical deployment of BS antennas. 
It is worth noting that in the position optimization scheme using only grid search, the performance improvement is much smaller compared to when Adam optimization is incorporated, whether in cases with a larger movable range or more constrained conditions. This highlights the crucial role of Adam in finding the optimal position for the antenna with high precision.
Additionally, we compare the three iteration cases when \( R = 0.04 \). When the antenna's movable range is very small in the horizontal dimension due to constraints in the horizontal direction, the performance gain (22\%) is lower compared to the case when \( R = 0.5 \). Therefore, in practical deployments, the movement radius of the circular trajectory should not be too small, as this would restrict the antenna's movable range and reduce optimization performance.

\begin{figure}
	\centering
	\includegraphics[width=0.98\linewidth]{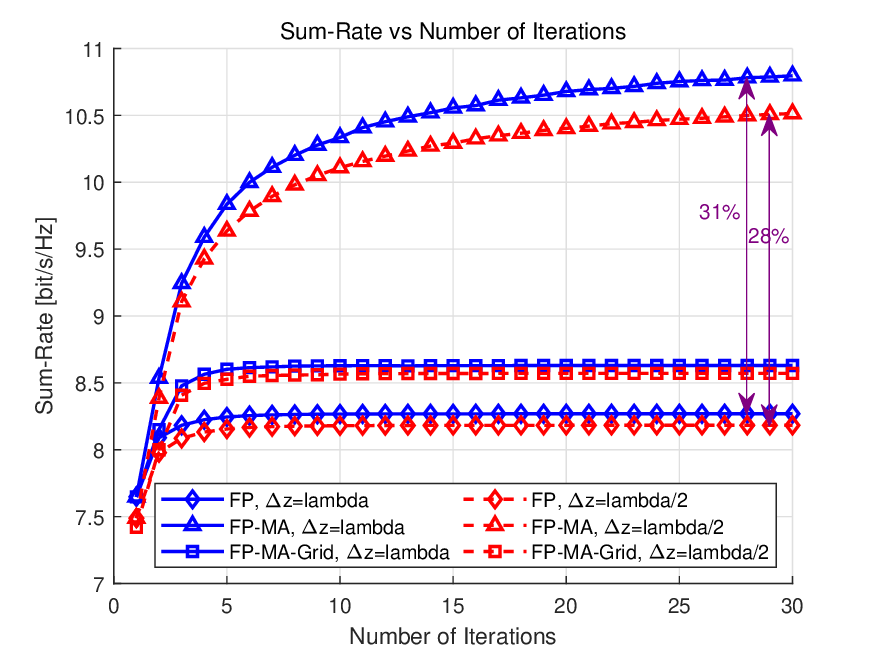}
	\caption{Sum-Rate vs. Iterations for Different Position Optimization Schemes and Vertical Movable Ranges}
	\label{fig_iterz}
\end{figure}

Fig. \ref{fig_iterz} shows the different movable ranges for the vertical dimension, comparing the performance of the three schemes.
When the initial height of each antenna layer is uniformly distributed with a spacing of \( \lambda_0/2 \), the movable range in the vertical dimension is smaller, and the optimization performance is more significantly affected by constraints. As a result, the performance improvement (28\%) is smaller compared to when the spacing is \( \lambda \). Therefore, in practical deployments, it is important to avoid having too small a spacing between layers. Similarly, it can be observed that the performance improvement from using only grid search is much smaller than that achieved by incorporating Adam optimization, highlighting the indispensable role of Adam in the position optimization process.

\begin{figure}
	\centering
	\includegraphics[width=0.98\linewidth]{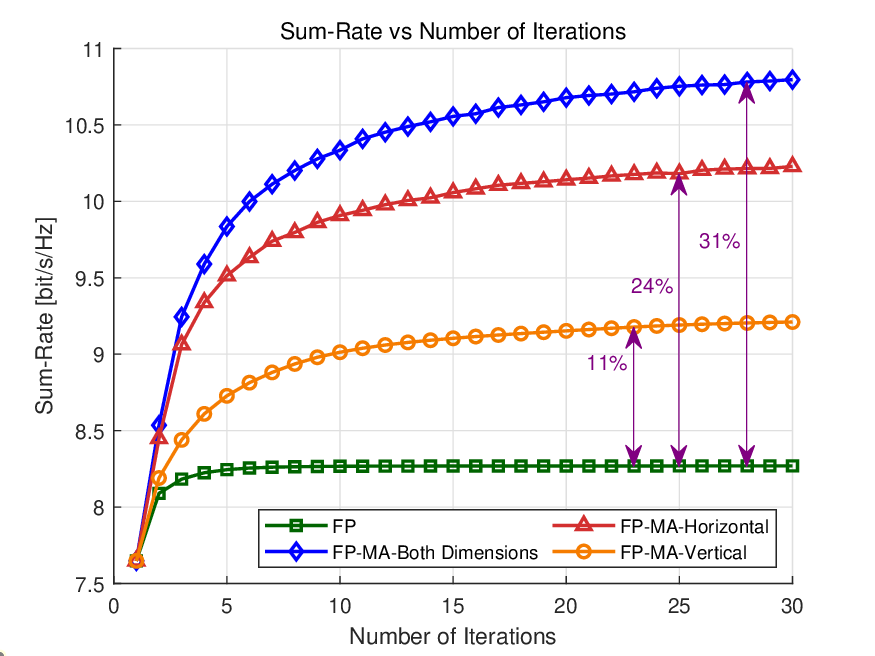}
	\caption{Sum-Rate vs. Iterations with Optimization in Different Dimensions}
	\label{fig_iter_demen}
\end{figure}

We consider the impact of optimization in different dimensions on the overall gain. We compare the case where only horizontal optimization \( \bm{\psi} \) is performed without optimizing \( \mathbf{z} \) in the vertical dimension, the case where only vertical optimization of \( \mathbf{z} \) is performed without horizontal optimization, and the case where optimization is conducted in both dimensions, as shown in Fig. \ref{fig_iter_demen}. 
The meanings of each curve are represented as follows:
\begin{itemize}
	\item \textbf{FP}: Represents the Fixed Antenna Position Based on the FP scheme without MAs.
	\item \textbf{FP-MA-Both Dimensions}: Represents optimization in both the horizontal and vertical dimensions.
	\item \textbf{FP-MA-Horizontal}: Represents optimization only in the horizontal dimension.
	\item \textbf{FP-MA-Vertical}: Represents optimization only in the vertical dimension.
\end{itemize}
All of these use CGS-Adam for optimization. The results show that the performance improvement from horizontal dimension optimization is much greater than that from vertical dimension optimization, indicating that antenna angle optimization in the horizontal dimension is more important.

\begin{figure}
	\centering
	\includegraphics[width=0.98\linewidth]{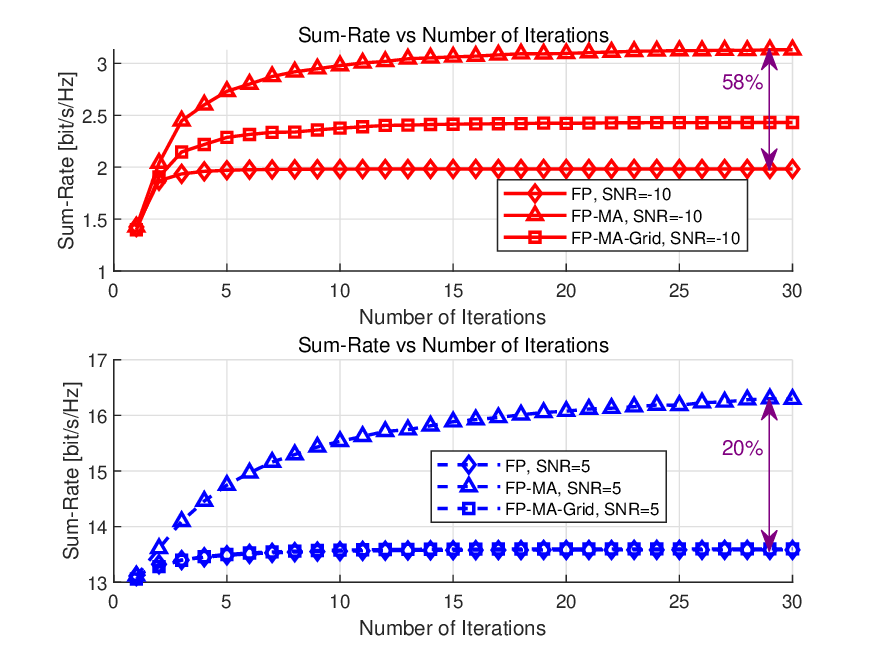}
	\caption{Sum-Rate vs. Iterations for Optimization at Different SNRs}
	\label{fig_iter_SNR}
\end{figure}

Fig. \ref{fig_iter_SNR} illustrates the convergence behavior of three schemes with increasing iteration counts under different Signal-to-Noise Ratios (SNRs). It simulates the case of SNR = -10, where the noise is significantly high and the data transmission quality is very poor, as well as SNR = 5, where the noise intensity is lower and the communication quality is better. The convergence of the three schemes is shown in both poor and favorable communication environments. It can be observed that, regardless of the communication environment, the MA scheme converges and improves. Furthermore, the performance of the CGS-Adam approach shows a higher improvement compared to the grid search scheme, with a performance enhancement of up to 58\% under high noise conditions.

\begin{figure}
	\centering
	\includegraphics[width=0.98\linewidth]{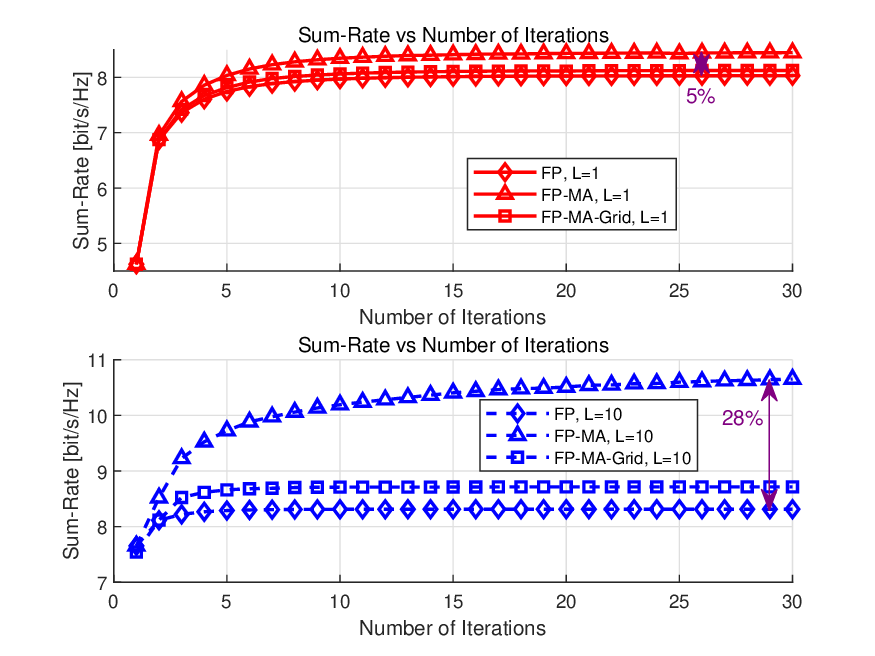}
	\caption{Sum-Rate vs. Iterations for Optimization at Different Path Counts \( L \)}
	\label{fig_iter_L}
\end{figure}

Similarly, we compare the iteration convergence of sum-rate at different path counts \( L \), as shown in Fig. \ref{fig_iter_L}. The convergence trends of the three schemes exhibit similar patterns at both low path counts \( L = 1 \) and high path counts \( L = 10 \). The sum-rate value at \( L = 10 \), where the number of paths is higher, is greater than that at \( L = 1 \), where the number of paths is smaller. Additionally, we observe that with an increased number of paths, the sum-rate gain is higher compared to the case with fewer paths, which is more applicable to MIMO systems.

\begin{figure}
	\centering
	\includegraphics[width=0.98\linewidth]{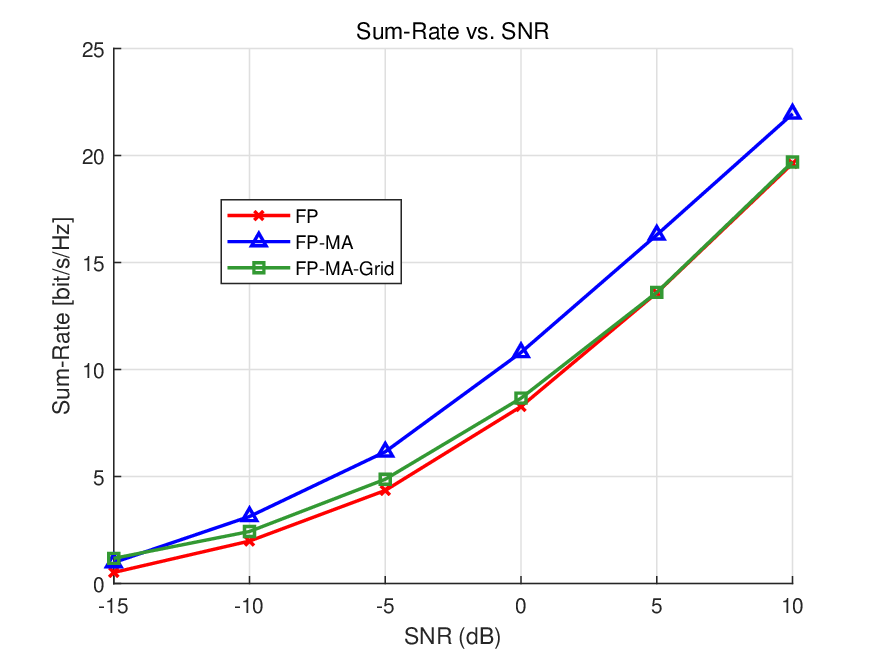}
	\caption{Sum-Rate vs. SNR}
	\label{fig_SNRrange}
\end{figure}

Fig. \ref{fig_SNRrange} shows the final convergence values of the three schemes under different SNR conditions. As can be observed, the final convergence results increase with the SNR, which aligns with our expectations. Furthermore, the trends of the three schemes remain consistent, with the CGS-Adam approach outperforming the traditional fixed-antenna scheme.

\begin{figure}
	\centering
	\includegraphics[width=0.98\linewidth]{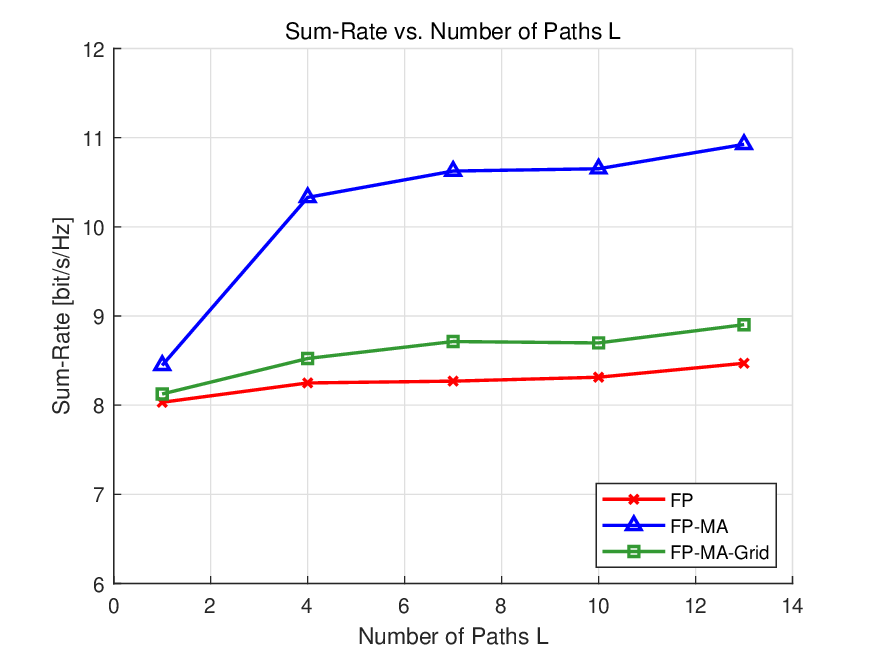}
	\caption{Sum-Rate vs. Path Count \( L \)}
	\label{fig_Lrange}
\end{figure}

The corresponding Fig. \ref{fig_Lrange} shows the final convergence behavior under different numbers of propagation paths \( L \). As observed, the CGS-Adam approach continues to exhibit the best performance improvement. As \( L \) increases, the CGS-Adam approach, represented by the FP-MA curve, leverages the advantage of multi-path propagation more effectively by adjusting the antenna positions, thereby improving reliability and capacity, and increasing the total sum-rate.

\begin{figure}
	\centering
	\includegraphics[width=0.98\linewidth]{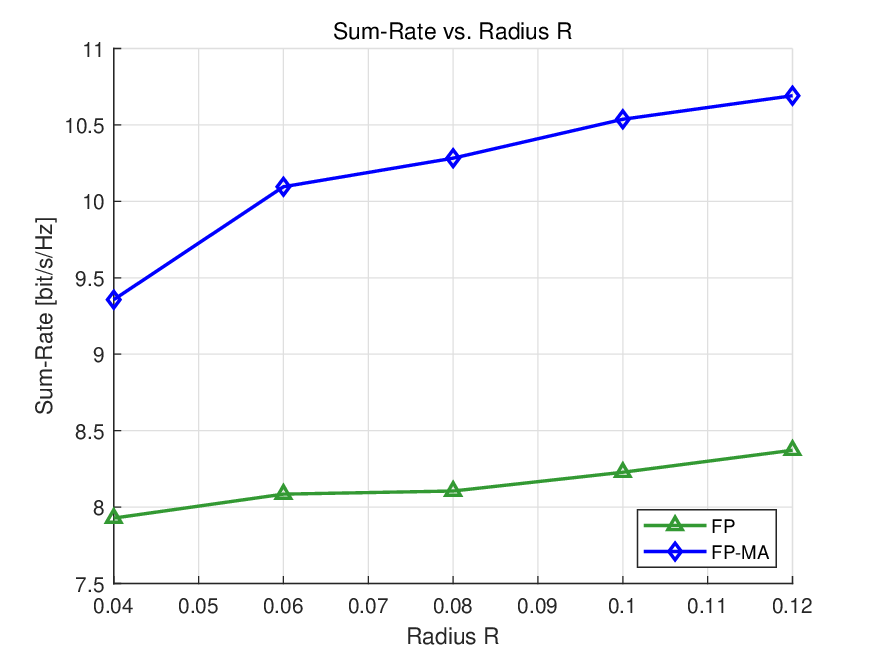}
	\caption{Sum-Rate vs. Radius \( R \)}
	\label{fig_Rrange}
\end{figure}

Fig. \ref{fig_Rrange} shows the variation in the final sum-rate convergence values under different radius \( R \). As observed, the CGS-Adam approach, which optimizes the antenna positions, consistently provides a higher gain compared to the fixed antenna optimization scheme. With an increase in \( R \), the available movement range in the horizontal dimension expands, leading to an increase in the sum-rate value under the CGS-Adam approache.

%
  
\section{Conclusions}\label{Con} 

In this paper, we have proposed an innovative optimization framework for wireless communication systems utilizing MAs integrated into FCLA. By leveraging the dynamic mobility of the FCLA, we optimize the performance of MU-MISO downlink systems. Our proposed system employs the FP algorithm to alternate between beamforming, antenna revolving angle, and heights, ensuring the efficient use of system resources and boosting communication rates.

We have proposed a novel constrained CGS-Adam algorithm for optimizing antenna positions while respecting spacing constraints. This algorithm enhances convergence speed and ensures optimal antenna positions, further improving system performance. Through extensive simulations, we have demonstrated that the integration of MAs significantly outperforms traditional fixed antenna optimization methods, achieving up to a 31\% performance improvement in general scenarios.


The use of FCLAs in wireless networks offers substantial advantages in terms of coverage, flexibility, and energy efficiency, making them a promising solution for the future 6G era. The real-time adaptability of MAs, in combination with advanced algorithmic strategies, can address the dynamic challenges posed by complex wireless environments, such as interference and signal fading. As the demand for ultra-reliable, high-speed, and ubiquitous connectivity continues to grow, the integration of FCLAs and MAs into communication systems is poised to play a pivotal role in advancing next-generation wireless networks.
 
\begin{appendices} 
\end{appendices}

\bibliographystyle{IEEEtran}
\bibliography{reference.bib}

\vspace{12pt}

\end{document}